\documentclass[%
 reprint,
superscriptaddress,
 amsmath,amssymb,
 aps,
]{revtex4-2}
\usepackage{graphicx}% Include figure files
\usepackage{dcolumn}% Align table columns on decimal point
\usepackage{bm}% bold math
\usepackage[version=3]{mhchem} %chemical fomulas
\usepackage{color}
\usepackage{ulem}

\begin{document}
\title{Resonant and Bound States of Charged Defects in Two-Dimensional Semiconductors}
\author{Martik Aghajanian}
\affiliation{Departments of Physics and Materials and the Thomas Young Centre for Theory and Simulation of Materials, Imperial 
College London, London, SW7 2AZ, UK.}
\author{Bruno Schuler}
\email{bschuler@lbl.gov}
\author{Katherine A. Cochrane}
\affiliation{Molecular Foundry, Lawrence Berkeley National Laboratory, Berkeley, California 94720, USA.}
\author{Jun-Ho Lee}
\affiliation{Molecular Foundry, Lawrence Berkeley National Laboratory, Berkeley, California 94720, USA.}
\affiliation{Department of Physics, University of California at Berkeley, Berkeley, California 94720, USA.}
\author{Christoph Kastl}
\affiliation{Molecular Foundry, Lawrence Berkeley National Laboratory, Berkeley, California 94720, USA.}
\affiliation{Walter-Schottky-Institut and Physik-Department, Technical University of Munich, Garching 85748, Germany.}
\author{Jeffrey B. Neaton}
\affiliation{Molecular Foundry, Lawrence Berkeley National Laboratory, Berkeley, California 94720, USA.}
\affiliation{Department of Physics, University of California at Berkeley, Berkeley, California 94720, USA.}
\affiliation{Kavli Energy Nanosciences Institute at Berkeley, Berkeley, California 94720, USA.}

\author{Alexander Weber-Bargioni}
\affiliation{Molecular Foundry, Lawrence Berkeley National Laboratory, Berkeley, California 94720, USA.}
\author{Arash A. Mostofi}
\author{Johannes Lischner}
\email{j.lischner@imperial.ac.uk}
\affiliation{Departments of Physics and Materials and the Thomas Young Centre for Theory and Simulation of Materials, Imperial 
College London, London, SW7 2AZ, UK.}
\date{\today}
\begin{abstract}
A detailed understanding of charged defects in two-dimensional semiconductors is needed for the development of ultrathin electronic devices. Here, we study negatively charged acceptor impurities in monolayer WS$_2$ using a combination of scanning tunnelling spectroscopy and large-scale atomistic electronic structure calculations. We observe several localized defect states of hydrogenic wave function character in the vicinity of the valence band edge. Some of these defect states are bound, while others are resonant. The resonant states result from the multi-valley valence band structure of WS$_2$, whereby localized states originating from the secondary valence band maximum at $\Gamma$ hybridize with continuum states from the primary valence band maximum at K/K$^{\prime}$. Resonant states have important consequences for electron transport as they can trap mobile carriers for several tens of picoseconds.
\end{abstract}
\maketitle

Impurity doping is the prime technology to control the electrical conductivity of semiconductors over several orders of magnitude~\cite{Sun2014, VandeWalle2011}. 
Dopants introduce additional charge carriers that bind to the impurity and can be excited into the delocalized conduction or valence band states. Through this ionization process, the density of mobile charge carriers in the material is increased. However, the Coulomb potential induced by the positively charged donor or negatively charged acceptor scatters mobile charge carriers and thereby reduces carrier mobility.

Importantly, the Coulomb potential of charged impurities is screened by the dielectric response of the host semiconductor. The resulting perturbation leads to the formation of localized states composed of Bloch states near the band edges~\cite{Bassani1974}. For bulk semiconductors, such as silicon or gallium arsenide, these defect states have hydrogenic wavefunctions~\cite{Xie2017, Zhang2013, Bastard1981, Karazhanov2003}. Because of the large dielectric constants and small effective masses of traditional bulk semiconductors, the wavefunctions extend over several nanometers and the corresponding binding energies are of the order of a few tenths of an electronvolt~\cite{kohn1957}.

Many semiconductors exhibit multiple conduction or valence band valleys, i.e., their band structure features multiple extrema in the vicinity of the band gap. Such an electronic structure can result in the formation of \textit{resonant impurity states}~\cite{Louie1976}. These states occur when a localized defect level originating from a secondary band extremum lies in the energy range of the continuum states of the primary band extremum~\cite{Karazhanov2003}. The hybridization with the continuum states leads to an energy shift and a broadening of the localized level into a resonance. Such resonant defect states play an important role in transport properties~\cite{Yu2018, ZhangQinyong2012} because a mobile charge carrier can be temporarily trapped by the impurity, which reduces the carrier mobility~\cite{Bassani1974}.  

Two-dimensional (2D) semiconductors, such as monolayer transition-metal dichalcogenides (TMDs), have been intensely studied over the last decade as promising candidates for nano-electronic components, optoelectronic devices, and quantum information applications~\cite{Kormanyos2014, Baugher2014}. Understanding the properties of charged defects in these materials is a crucial step toward developing a viable route for impurity doping analogous to traditional (3D) semiconductors. A key difference between 2D and 3D semiconductors is their dielectric response to a charged perturbation. Specifically, the screened potential due to a point charge in 2D is not well-described by the bare Coulomb potential divided by a dielectric constant as in bulk semiconductors. Instead, in 2D semiconductors the dielectric response is relatively weak and highly anisotropic, which gives rise to strongly bound excitons that dominate their optical response~\cite{mak2010,qiu2013}. Unconventional screening of charged defects in monolayer TMDs with multi-valley band structure and large spin-orbit effects can also give rise to bound defect states with unusual properties, e.g., the prediction that the most strongly bound acceptor state switches from being of K/K$^{\prime}$-valley character to $\Gamma$-valley character at a critical value of the defect charge~\cite{Aghajanian2018}.

In this Letter, we study the electronic properties of charged impurities in monolayer WS$_2$ on a graphene/SiC substrate using scanning tunneling microscopy / spectroscopy (STM/STS) and atomistic quantum-mechanical calculations. Using STS, we find that negatively charged impurities give rise to strong upwards band bending and a series of localized defect states in the vicinity of the valence band edge with $s$- and $p$-like character. Our calculations suggest that the $s$-like states are bound defect states originating from either the K/K$^{\prime}$ or $\Gamma$ valleys, while the $p$-like states are resonant defect states which result from the hybridization of localized states from the $\Gamma$ valley with continuum states from the K/K$^{\prime}$ valleys. Finally, we discuss the relevance of these findings for transport devices.

\begin{figure*}
    \centering
    \includegraphics[width=\textwidth]{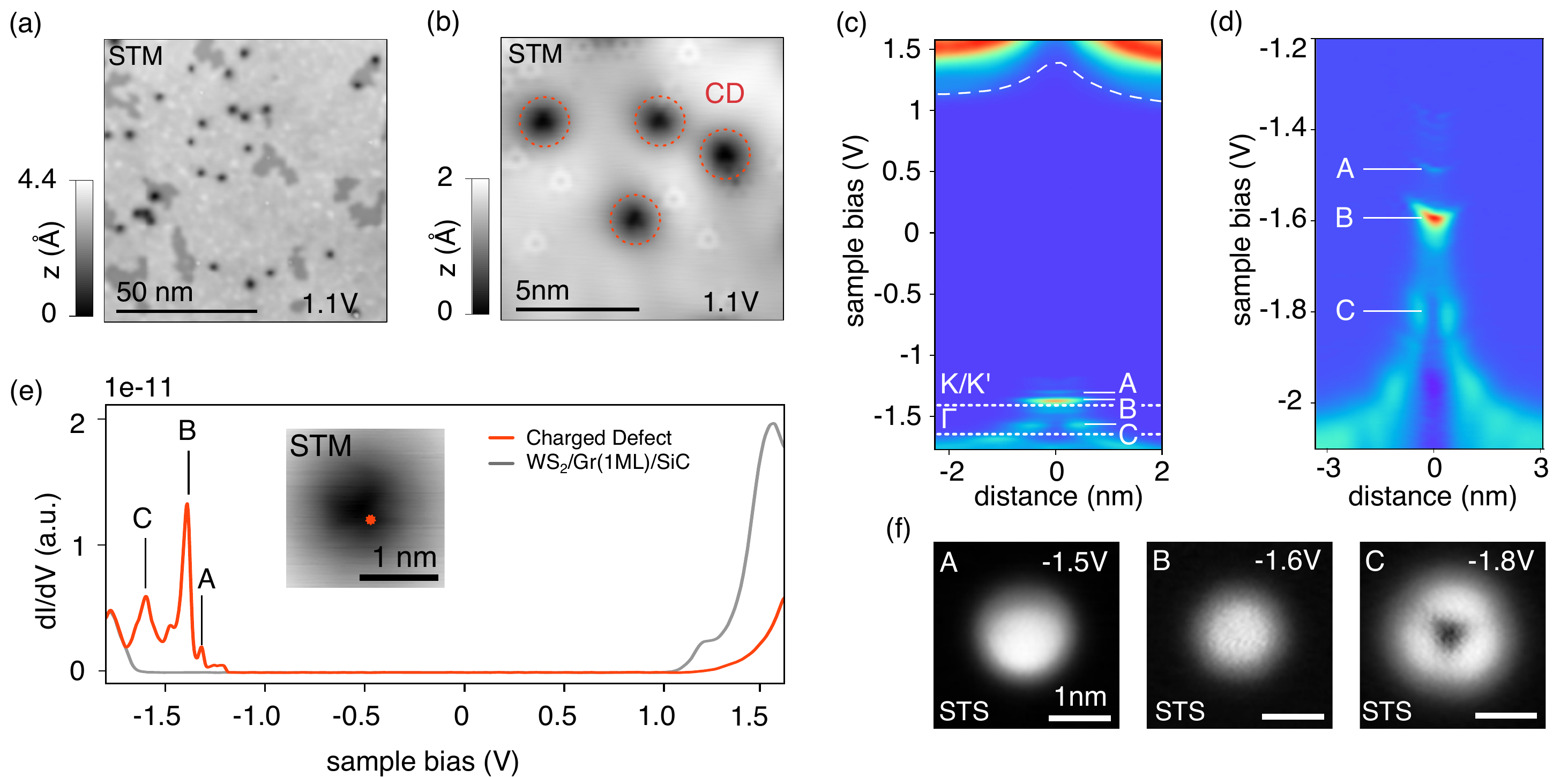}
    \caption{(a) STM topographic overview of the \ce{WS2}/graphene/SiC surface, with an applied sample bias, $V_b$ = 1.1\,V. (b) STM image showing defects in \ce{WS2}. The charged defects (CDs) are indicated with a dashed red circle. (c,d) STS across a charged defect showing band bending and localized defect states labeled A, B, and C. In (c) the upward bending of the conduction band due to the negatively charged defect is indicated with a dashed white line, and the onset of the K/K$^\prime$ and $\Gamma$ valleys of pristine \ce{WS2} are indicated with white dotted lines.  Note that the spectra in (c) and (d) were taken on different defects. (e) Point STS of a CD, the spectrum position is marked in the inset.  (f) Spatially resolved constant height $\mathrm{d}I/\mathrm{d}V$ maps of the defect corresponding to the electronic states A, B, and C.}
    \label{fig:stm}
\end{figure*}

STM/STS measurements were performed under ultrahigh vacuum ($<2\times10^{-10}$\,mbar) at low temperature ($\sim$6 K). \ce{WS2} samples were grown \textit{ex situ} by chemical vapor deposition (CVD) on epitaxial graphene on silicon carbide substrates \cite{kastl2017important} and subsequent annealing in vacuum at $\sim$200$^{\circ}$C. A representative overview image of the surface is shown in Fig.~\ref{fig:stm}(a). In our samples, we find various point defects, including oxygen-in-sulfur substitutions (O\textsubscript{S}), molybdenum-in-tungsten substitutions (Mo\textsubscript{W}), and charged defects [see Fig.~\ref{fig:stm}(b)], which we refer to as ``CDs" and have identified in the figure with red dashed circles. \cite{schuler2019overview} They are among the most abundant point defects in our samples but their density varies considerably between different CVD preparations. Similar STM contrast has been observed in CVD-grown MoS$_2$~\cite{liu2016point}, MOCVD- and MBE-grown WSe$_2$~\cite{lin2018realizing,le2018band}, MBE-grown MoSe$_2$~\cite{le2018band} and natural bulk MoS$_2$(0001)~\cite{addou2015surface}. Some of these reports suggest that the observed defect might be charged~\cite{addou2015surface,le2018band,schuler2019overview} but their chemical origin remains unclear.

Non-contact atomic force microscopic (nc-AFM) imaging with a carbon monoxide (CO) functionalized tip~\cite{Gross2009a} indicates that the charged defect is a chalcogen substitution~\cite{schuler2019overview}. The chemical origin of the CDs is yet unknown. Possible candidates include a CH or N substitution at the S site~\cite{schuler2019overview}. These substitutions introduce unoccupied defect states close the valence band edge of WS$_2$. When the TMD is placed on the metallic graphene/SiC substrate, the Fermi level is pinned in the upper half of the WS$_2$ band gap and the unoccupied defect states become filled due to charge transfer from the graphene~\cite{ulstrup2019nanoscale}. As the hybridization between the WS$_2$ states and the substrate is weak~\cite{schuler2018large}, the impurity charge is an integer multiple of the electron charge. 

\begin{figure}
    \centering
    \includegraphics[width=0.5\textwidth]{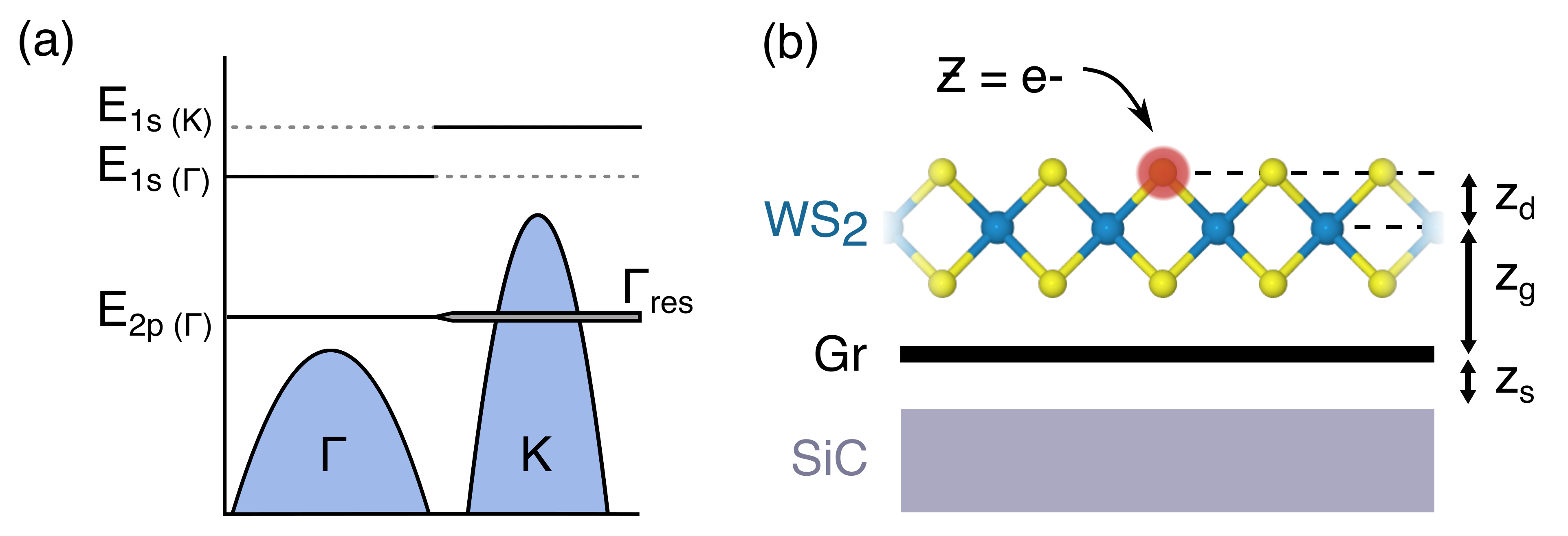}
    \caption{(a) Schematic of the electronic structure of a negatively charged defect in WS$_2$. Shown are the primary valence band maximum at K (a second degenerate maximum exists at K$^{\prime}$) and the secondary maximum at $\Gamma$ as well as the bound and resonant defect states. The states labelled 1$s$(K), 1$s$($\Gamma$) and 2$p$($\Gamma$) correspond to the experimental features A, B and C in Fig.~\ref{fig:stm}, respectively. (b) Schematic of the system consisting of a WS$_2$ layer on graphene on a silicon carbide substrate with a substitutional impurity with charge $Z$. Distances $z_\mathrm{d}$, $z_\mathrm{g}$, and $z_\mathrm{s}$ are indicated (see SM for details about the determination of these parameters~\cite{SI}).}
    \label{fig:setup}
\end{figure}

In constant-current STM imaging, CDs are observed as large depressions at positive sample bias and protrusions at negative sample bias. This is an electronic effect: a negative charge located at the CD site results in local upwards band bending. As shown in the $\mathrm{d}I/\mathrm{d}V$ spectra taken across a CD in Fig.~\ref{fig:stm}(c), electrons are pushed to higher energies in the vicinity of the defect, leading to a position-dependent onset of the conduction band minimum (CBM), indicated with a white dashed line. The valence band of WS$_2$ exhibits two primary maxima at the K and K$^{\prime}$ points of the Brillouin zone and a secondary maximum at the $\Gamma$ point [see Fig.~\ref{fig:setup}(a)]. The energies of these band extrema measured far away from the defect are indicated by white dotted lines in Fig.~\ref{fig:stm}(c). It should be noted that states originating from the K/K$^\prime$ point give only a weak STS signal. This is due to two factors: their relatively small effective mass results in a small local density of states (LDOS) and their large crystal momentum leads to a small tunnelling matrix element~\cite{zhang2008}. In contrast, the combination of a larger effective mass and a smaller crystal momentum at $\Gamma$ results in a stronger STS signal.

While there are no observed defect states near the CBM, several defect resonances are observed at negative bias in the vicinity of the valence band edge [see Figs.~\ref{fig:stm}(c,d)]. In Fig.~\ref{fig:stm}(e), we have labeled the three dominant resonances as ``A", ``B", and ``C", respectively. In addition, there are four very faint resonances observed just above state A [see the Supplementary Materials (SM)~\cite{SI}]. We find that the localized defect states A and B lie on average 175\,mV and 95\,mV, respectively, above the onset of the K/K$^\prime$ valence band edge (see Table~\ref{tab:binding}), indicating that they are bound states. Resonance C lies $130$\,mV below the K/K$^\prime$ valence band edge, but $110$\,mV above the onset of the $\Gamma$ valence band edge, suggesting that it is a resonant state, i.e., a localized state resonant with the continuum states from the K/K$^{\prime}$ valleys.

\begin{table}[!h]
\centering
\setlength{\tabcolsep}{1.25em}
\def\arraystretch{1.5}
%\begin{tabular}{p{1cm}p{2.5cm}p{2.5cm}|p{2.5cm}p{2.5cm}}
\begin{tabular}{ccc|cc}
  & \multicolumn{2}{c}{Experiment} & \multicolumn{2}{c}{Theory}\\
  \cline{2-3} \cline{4-5}
 & K & $\Gamma$ & K & $\Gamma$ \\ \hline \hline
\multicolumn{1}{l|}{A/meV} & \multicolumn{1}{r}{$+175$} & \multicolumn{1}{r|}{$+415$} & \multicolumn{1}{r}{$+95$} & \multicolumn{1}{r}{$+355$}\\ \hline
\multicolumn{1}{l|}{B/meV} & \multicolumn{1}{r}{$+95$} & \multicolumn{1}{r|}{$+335$} & \multicolumn{1}{r}{$+25$} & \multicolumn{1}{r}{$+285$}  \\ \hline
\multicolumn{1}{l|}{C/meV} & \multicolumn{1}{r}{$-130$} & \multicolumn{1}{r|}{$+110$} & \multicolumn{1}{r}{$-180$} & \multicolumn{1}{r}{$+80$} \\ \hline
\end{tabular}
\caption{Comparison of experimentally determined and theoretically calculated binding energies of three observed resonances: A, B, and C. Values are listed with respect to the K and $\Gamma$ valence band edges. Positive binding energies indicate a lower energy with respect to the band edge.}
\label{tab:binding}
\end{table}

Spatially resolved $\mathrm{d}I/\mathrm{d}V$ maps of the states A, B, and C are shown in Fig.~\ref{fig:stm}(f). States A and B have a spherical shape reminiscent of an $s$-type orbital, whereas state C has three lobes and a node at its center, indicative of a trigonally warped $2p$-type orbital~\cite{Aghajanian2018}.
With a lateral dimension of about $1.5-2$\,nm, the localized defect states are closely confined, as compared to the Rydberg states recently reported for ionized defects in bulk black phosphorus which extend more than ten nanometers~\cite{qiu2017resolving}. The smaller size of the acceptor states in our experiment is a consequence of the different environmental screening. Specifically, the metallic screening from the doped graphene substrate leads to more localized defect states compared to the states one would obtain for an insulating substrate.

To understand the experimental observations, we carried out theoretical calculations of acceptor impurities in WS$_2$ using a recently developed approach that allows the simulation of very large supercells that are required to accurately describe shallow defect states of charged impurities~\cite{Aghajanian2018}. In particular, we first calculated the screened potential of the defect, which is modelled as a point charge with $Z=-1$ located $z_\mathrm{d}=2.4\text{ \AA}$ above the plane of the tungsten atoms, see Fig.~\ref{fig:setup}(b). To describe the screening of the defect charge, the dielectric function of the system is computed using the random phase approximation (RPA). For this, we compute the polarizability as the sum of contributions from the WS$_2$, the doped graphene and the SiC substrate (see SM for details~\cite{SI}). 

To calculate the LDOS of a single charged impurity in WS$_2$, we carry out large-scale tight-binding calculations using a three-band tight-binding model~\cite{liu15} with the screened defect potential included as an on-site energy~\cite{Aghajanian2018}. This allows us to obtain converged results for a $45\times 45$ WS$_2$ supercell. This supercell, which would be extremely challenging to model with \textit{ab initio} techniques, is sufficiently large to capture the decay of the shallow impurity states and obtain accurate defect state energies. While this approach allows us to describe states induced by the long-ranged screened defect potential, the model does not capture short-ranged chemical interactions that could give rise to additional defect states.

\begin{figure*}[t!]
\centering
\includegraphics[width=\textwidth]{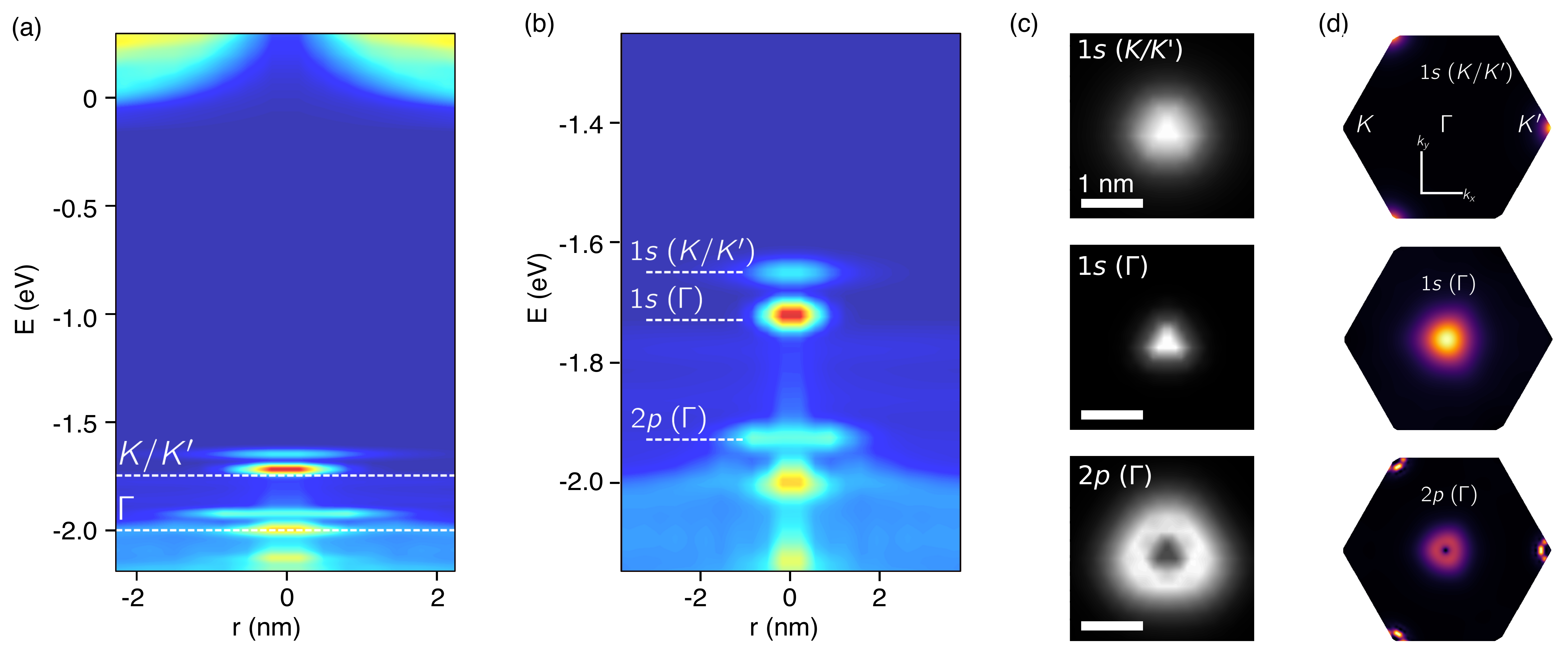}
\caption{(a), (b) Calculated local density of states of a negatively charged defect in WS$_2$ on graphene/SiC substrate. In (a), the onset of the K/K$^{\prime}$ and $\Gamma$ valleys are indicated by dashed lines. In (b), the defect states are labeled by their hydrogenic quantum numbers and valley. (c)
Simulated STS images of defect states. (d) Projections of defect wavefunctions onto unperturbed WS$_2$ states.}
\label{fig:sim}
\end{figure*}

Figures~\ref{fig:sim}(a) and (b) show the calculated LDOS of a single acceptor impurity in WS$_2$. In agreement with the measured $\mathrm{d}I/\mathrm{d}V$ map [Fig.~\ref{fig:stm}(c)], we observe a significant defect-induced upward bending of the conduction band edge. In the vicinity of the valence band edge, a series of defect states are found. Fig.~\ref{fig:sim}(c) shows the simulated STM images of the three most strongly bound defect states that give rise to significant peaks in the LDOS (see SM for details of STM simulations~\cite{SI}). The two most strongly bound states appear similar to experimental states A and B as circular bright spots with a similar size [Fig.~\ref{fig:stm}(f)]. The binding energies with respect to the valence band maximum at K/K$^\prime$ of these states are 95\,meV and 25\,meV, respectively. While the absolute binding energies are somewhat smaller than the experimental values, see Table~\ref{tab:binding}, their difference (70\,meV) is in good agreement with experiment (80\,meV). As these states exhibit symmetries similar to the eigenstates of the two-dimensional hydrogen atom, we label them as $1s$ states. Interestingly, we find that the more strongly bound $1s$ state is less localized than the other $1s$ state. At lower binding energies, a $2p$-like defect state is found that is characterized by a node at its center. This state resembles the state C in Fig.~\ref{fig:stm}(f). The binding energy of this states is $-180$\,meV indicating that it lies between $\Gamma$ and K/K$^\prime$. Again, we find that the binding energy difference between this state and the less strongly bound $1s$ state is in very good agreement with experiment (theory: 205\,meV, experiment: 225\,meV). 

To understand which band extrema these defect states originate from, we decompose the defect wave functions into contributions from unperturbed WS$_2$ states. Fig.~\ref{fig:sim}(d) shows that the most strongly bound defect state originates from the K valley (a second degenerate state with opposite spin originates from the K$^{\prime}$ valley) and is therefore labelled as $1s$ (K), while the other $1s$ state originates from the $\Gamma$ valley. As the effective mass of the $\Gamma$ valley is larger than that of the K/K$^{\prime}$ valleys, the $1s$($\Gamma$) state is more localized despite being less strongly bound [recall that for the 2D hydrogen atom $\psi_{1s}(r) \propto \exp(-2m e^2r/\hbar^2)$ indicating that a larger electron mass $m$ results in a more localized state]. The analysis of the $2p$ resonance reveals that this impurity level mostly originates from unperturbed states in the vicinity of $\Gamma$, but also has contributions from a few states near K (note that mixing only occurs between states with the same spin preventing any hybridization with states from K$^{\prime}$). The valley hybridization is a consequence of the resonant nature of the $2p$ defect state and can be understood within the framework of effective mass theory~\cite{Bassani1974}. In this approach, one first neglects the coupling between different valleys and obtains a set of bound defect states for each valley. When the coupling between valleys is turned on, the bound states from the secondary valence band extremum at $\Gamma$ that lie in the same energy range as continuum states from the K and K$^{\prime}$ valleys can mix with the continuum states. The resulting hybridized states can be expressed as
\begin{align}
    \psi_{2p(\Gamma)}(\mathbf{r}) &= c_{2p(\Gamma)}\phi_{2p(\Gamma)}(\mathbf{r}) \nonumber \\
    &\quad + \sum_{n\in\{ K,K^{\prime}\}}\int_{\Omega_n}\mathrm{d}\mathbf{k}\;c_n(\mathbf{k})\phi_{n\mathbf{k}}(\mathbf{r}),
\end{align}
where $\phi_{2p(\Gamma)}$ and $\phi_{n\mathbf{k}}$ denote the bound and continuum states obtained for decoupled valleys, respectively, (with $n$ labelling the valley), $c_{2p(\Gamma)}$ and $c_n(\mathbf{k})$ are complex coefficients, and $\Omega_n$ denotes the k-space region corresponding to the $n^{\mathrm{th}}$ valley.

Finally, we discuss the effect of resonant impurities on electron transport. If a hole near one of the primary band extrema (i.e., in the K or K$^{\prime}$ valley) in an external electric field reaches the energy of the resonant impurity state $E_{2p(\Gamma)}$, it does not participate in conduction for a time $\tau_{\text{res}}=\hbar\Gamma^{-1}_{\text{res}}$. Here, $\Gamma_{\text{res}}$ denotes the width of the resonant impurity level given by
\begin{equation}
\Gamma_{\text{res}}=\pi 
\sum_{n\in\{K,K^{\prime}\}}\int_{\Omega_n} d\mathbf{k} |V_{2p(\Gamma),n\mathbf{k}}|^2
\delta(E_{2p(\Gamma)}-\epsilon_{n\mathbf{k}}),
\end{equation}
where $\epsilon_{n\mathbf{k}}$ denotes the energy of a continuum state with crystal momentum $\mathbf{k}$ in valley $n$ and $V_{2p(\Gamma),n\mathbf{k}}$ is the matrix element of the Bloch-transformed screened impurity potential between $\phi_{2p(\Gamma)}(\mathbf{r})$ and $\phi_{n\mathbf{k}}(\mathbf{r})$~\cite{Bassani1974}. Evaluating this expression with our first-principles defect potential and tight-binding wave functions yields $\Gamma_{\text{res}}=0.02$~meV and $\tau_{\text{res}}=30$~ps. This result shows that resonant defect states exhibit extremely small linewidths corresponding to long trapping times. Thus, we predict that resonant states should lead to a drastic reduction in conductivity when the Fermi level reaches the resonant impurity level. Note that in a WS$_2$ transport device, graphene could not be used as substrate as most carriers would propagate through the graphene. Without graphene, the screening of the impurity charge would be reduced, resulting in a resonant impurity level that lies closer to the VBM at K/K$^\prime$. The trapping time, however, would not change significantly as a consequence of the flat density of state of WS$_2$ near the VBM.

In this Letter, we have studied the localized acceptor states associated with negatively charged point defects in WS$_2$ using scanning tunneling microscopy and spectroscopy. We identify these states as resonant and bound hydrogenic states of the screened Coulomb potential. Our large-scale atomistic electronic structure calculations reproduce the band bending and binding energies  of the defect states as well as their wave function shapes and sizes as observed in STS measurements. The hydrogenic states exhibit $s$- and $p$-type character and originate from both the K/K$^{\prime}$ and $\Gamma$ valence band valleys. Despite the hybridization with the primary valence bands around K and K$^{\prime}$, the resonant state emerging from $\Gamma$ is clearly resolved. We further predict that resonant states can trap mobile carriers for up to 30 picoseconds. Our findings highlight the importance of a detailed understanding of the effect of charged defects for the development of novel nano-electronic devices based on two-dimensional semiconductors. 

\textit{Acknowledgments:} This work was supported through a studentship in the Centre for Doctoral Training on Theory and Simulation of Materials at Imperial College London funded by the EPSRC (EP/L015579/1). We acknowledge the Thomas 
Young Centre under grant number TYC-101. Via J.L.'s membership of the UK's HEC Materials Chemistry Consortium, which is funded by EPSRC (EP/L000202, EP/R029431), this work used the ARCHER UK National Supercomputing Service (http://www.archer.ac.uk). The work performed at the Molecular Foundry was supported by the Office of Science, Office of Basic Energy Sciences, of the U.S. Department of Energy under Contract No. DE-AC02-05CH11231. B.S. appreciates support from the Swiss National Science Foundation under project number P2SKP2\_171770. A.W.-B. was supported by the U.S. Department of Energy Early Career Award. J.-H.L. and J.B.N. were supported by the Air Force Office of Scientific Research Hybrid Materials MURI under award number FA9550-18-1-0480. Computational resources for performing the ab initio calculations were provided by the National Energy Research Scientific Computing Center and the Molecular Foundry, DOE Office of Science User Facilities supported by the Office of Science of the U.S. Department of Energy under Contract No. DE-AC02-05CH11231.

\bibliography{ws2defect}
\end{document}

\title{Supplemental Material:\texorpdfstring{\\}{} Resonant and Bound States of Charged Defects in Two-Dimensional Semiconductors}

\author{Martik Aghajanian}
\affiliation{Departments of Physics and Materials and the Thomas Young Centre for Theory and Simulation of Materials, Imperial 
College London, London, SW7 2AZ, UK.}
\author{Bruno Schuler}
\email{bschuler@lbl.gov}
\author{Katherine A. Cochrane}
\affiliation{Molecular Foundry, Lawrence Berkeley National Laboratory, Berkeley, California 94720, USA.}
\author{Jun-Ho Lee}
\affiliation{Molecular Foundry, Lawrence Berkeley National Laboratory, Berkeley, California 94720, USA.}
\affiliation{Department of Physics, University of California at Berkeley, Berkeley, California 94720, USA.}
\author{Christoph Kastl}
\affiliation{Molecular Foundry, Lawrence Berkeley National Laboratory, Berkeley, California 94720, USA.}
\affiliation{Walter-Schottky-Institut and Physik-Department, Technical University of Munich, Garching 85748, Germany.}
\author{Jeffrey B. Neaton}
\affiliation{Molecular Foundry, Lawrence Berkeley National Laboratory, Berkeley, California 94720, USA.}
\affiliation{Department of Physics, University of California at Berkeley, Berkeley, California 94720, USA.}
\affiliation{Kavli Energy Nanosciences Institute at Berkeley, Berkeley, California 94720, USA.}

\author{Alexander Weber-Bargioni}
\affiliation{Molecular Foundry, Lawrence Berkeley National Laboratory, Berkeley, California 94720, USA.}
\author{Arash A. Mostofi}
\author{Johannes Lischner}
\email{j.lischner@imperial.ac.uk}
\affiliation{Departments of Physics and Materials and the Thomas Young Centre for Theory and Simulation of Materials, Imperial 
College London, London, SW7 2AZ, UK.}

\maketitle 

\tableofcontents
\newpage
\section{Calculation of the screened defect potential}
The screened defect potential is obtained by dividing the bare Coulomb potential by the dielectric function in the random phase approximation (RPA) given by 
\begin{equation}
    \varepsilon_{\text{RPA}}(q)= -v(q)\left[\chi_{\text{TMD}}(q)+e^{-z_{\mathrm{g}}q}\chi_{\text{Gr}}(q)\right]
    + \eta(q), 
    \label{eqn:diel}
\end{equation}
where $v(q)=2\pi e^2/q$ denotes the 2D Fourier transform of the Coulomb interaction. The dielectric function of Eq.\eqref{eqn:diel} is a sum of three contributions. The first term on the right hand side describes screening from the TMD, with $\chi_{\text{TMD}}(q)$ denoting the polarizability of WS$_2$ obtained from first-principles calculations based on the Adler-Wiser formula~\cite{Qiu2016}.  The second term captures the screening from the doped graphene monolayer that sits a distance $z_{\mathrm{g}}=4.9\text{ \AA }$ beneath the TMD (measured from the plane of the transition metal atoms). The value of $z_\mathrm{g}$ was obtained from density-functional theory (DFT) calculations including van der Waals interactions. Specifically, $\chi_{\text{Gr}}(q)$ is the RPA polarizability of doped graphene given by~\cite{Wunsch_2006}
\begin{align}
\chi_{\text{Gr}}(q) &= -\frac{2k_\text{F}}{\pi\hbar v_\text{F}}\left[1-\Theta(q-2k_\text{F})\Lambda(q)\right], \\
\Lambda(q)& = \frac{1}{2}\sqrt{1-\left(\frac{2k_\text{F}}{q}\right)^2}-\frac{q}{4k_\text{F}}\text{cos}^{-1}\left(\frac{2k_\text{F}}{q}\right),
\end{align}
where $k_\text{F}$ and $v_\text{F}$ denote the Fermi wavevector and Fermi velocity of the doped graphene, respectively. We use $v_\text{F}=1.15\times10^{6}$ ms$^{-1}$~\cite{Hwang2012} and $k_\text{F}=0.059\text{ \AA}^{-1}$~\cite{Emtsev2009}. 
Finally, the last term in Eq.~\eqref{eqn:diel} describes the screening by the SiC substrate located a distance $z_{\mathrm{s}}=1.7\text{ \AA }$ beneath the graphene layer~\cite{Yoon2013}. An image charge model yields 
\begin{equation}
    \eta(q) = \frac{\varepsilon_{\text{SiC}}+1}{\varepsilon_{\text{SiC}}+1-(\varepsilon_{\text{SiC}}-1)e^{-2q(z_\mathrm{s}+z_\mathrm{g})}},
\end{equation}
with $\varepsilon_{\text{SiC}}=9.7$ denoting the bulk dielectric constant of SiC~\cite{Lyle1970}.

The defect potential in real space is obtained via a Hankel transform~\cite{Aghajanian2018} 
\begin{equation}
    V(r) = Ze^2\int\mathrm{d}q\;\frac{e^{-qz_\mathrm{d}}J_0(qr)}{\varepsilon_{\text{RPA}}(q)},
\end{equation}
where $Z$ and $z_\mathrm{d}$ denote the defect charge and the distance of the defect above the transition-metal plane, respectively. We assume a defect charge of $Z=-1$ and use $z_\mathrm{d}=2.4\text{ \AA}$. We note that $z_\mathrm{d}$ should not be interpreted as the actual defect height. Instead, this parameter regularizes the Coulomb interaction at short distances and reflects the central cell correction that is required to capture chemical effects close to the impurity~\cite{Bassani1974}. We have carried out calculations for different values $z_\mathrm{d}$ and determined that 2\,{\AA} gives the best agreement with experiment for the LDOS (see SM~\cite{SI}).

To obtain the dielectric function of WS$_2$ within the random-phase approximation, we first perform density-functional theory (DFT) calculations of a structurally relaxed ($a_{\text{lat}}=3.19\text{ \AA}$) monolayer of WS$_2$ using the PBE exchange-correlation functional with a plane-wave cut-off of 80~Ry and a $12\times12\times 1$ $k$-point mesh as implemented in the Quantum Espresso software package~\cite{giannozzi2009}. To prevent interactions between periodically repeated monolayers, we employ a separation in the out-of-plane direction of $L_z=18\text{ \AA}$. We then calculate 2900 conduction band states on a finer $30\times30\times 1$ $k$-point mesh, and utilise the BerkeleyGW package~\cite{deslippe2012} to determine the inverse dielectric matrix $\varepsilon^{-1}_{\mathbf{GG'}}(\mathbf{q})$, using a truncated reciprocal-space Coulomb interaction $V_{\text{trunc}}(\mathbf{q})$~\cite{ismail2006}. From this we extract an effective two-dimensional dielectric function of the TMD monolayer using \cite{Qiu2016}
\begin{equation}
    \varepsilon^{-1}_{\text{TMD}}(\mathbf{q}) = \frac{q}{2\pi e^2L_z}\sum_{\mathbf{G}_z\mathbf{G}^{\prime}_{z}}W_{\mathbf{G_zG^{\prime}_{z}}}(\mathbf{q}), 
\end{equation}
where $W_{\mathbf{G_zG^{\prime}_{z}}}(\mathbf{q})=\varepsilon^{-1}_{\mathbf{G_zG^{\prime}_{z}}}(\mathbf{q})V_{\text{trunc}}(\mathbf{q+G^{\prime}_{z}})$ denotes the screened interaction. We define the polarization function $\chi_{\text{TMD}}(\mathbf{q})$ via the expression $ \varepsilon_{\text{TMD}}(\mathbf{q})=1-v(\mathbf{q})\chi_{\text{TMD}}(\mathbf{q})$, where $v(\mathbf{q})$ is the 2D Fourier transform of the bare Coulomb potential.

\section{Determination of $z_\mathrm{g}$}

To obtain a value for $z_\mathrm{g}$, the distance between the WS$_2$ and the graphene, we carried out DFT calculations including van der Waals (vdW) interactions. Specifically, we studied a supercell that contains $4\times4=16$ unit cells of WS$_2$ on top of $5\times5=25$ unit cells of graphene. A $5\times5\times1$ $k$-point mesh was used. To deal with the lattice mismatch between graphene and WS$_2$, we applied 2.13\% in-plane compressive strain to the WS$_2$. In order to take vdW interactions into account, we used the rev-vdW-DF2 scheme~\cite{Hamada2014} that accurately describes the geometry of layered materials~\cite{Kim2017Origins}. To determine the value of $z_\mathrm{g}$, we averaged over the z-coordinates of the C and W atoms in the supercell.

\section{Determination of $z_\text{d}$}
\begin{figure}[!h]
    \centering
    \includegraphics[width=0.5\textwidth]{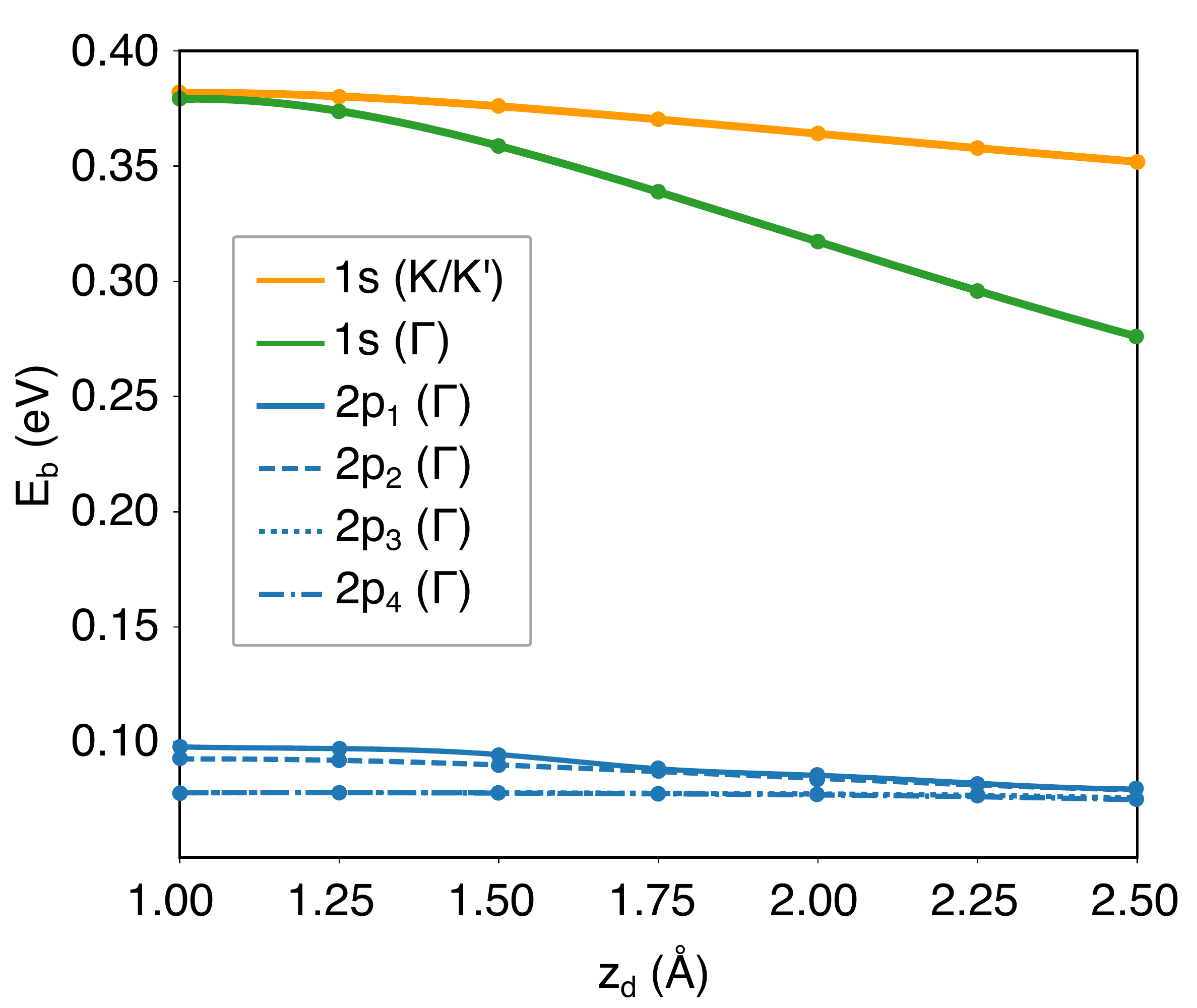}
    \caption{Dependence of the binding energy (measured with respect to the secondary valence band maximum at $\Gamma$) of various defect states on the parameter $z_\mathrm{d}$.}
    \label{fig:vary}
\end{figure}

We have carried out calculations of the defect binding energies for various values of the parameter $z_\mathrm{d}$, which formally can be interpreted as the height of the defect above the plane of the transition metal atoms. Our results are shown in Fig.~\ref{fig:vary}. As $z_\mathrm{d}$ increases, the potential experienced by the electrons in the WS$_2$ is weaker resulting in smaller binding energies. The $1s$ states are more sensitive to changes in $z_\mathrm{d}$ than the $2p$ states as the most significant changes to the potential occur in the WS$_2$ region underneath the defect where the $2p$ states have a node (note that the hybridization of the localized $2p$ state from the $\Gamma$ valley with the continuum states from the K/K$^\prime$ valleys gives rise to multiple mixed states with are labelled by $2p_n(\Gamma)$ in Fig.~\ref{fig:vary}). As a consequence of the larger effective mass of the $\Gamma$ valley, the $1s(\Gamma)$ state is more localized than the $1s$(K/K$^\prime$) state and therefore decreases more quickly as $z_\mathrm{d}$ is increased. For a value of $z_\mathrm{d}=2\text{ \AA}$, the separation between $1s$(K/K$^{\prime}$) and $1s$($\Gamma$) is about 0.1~eV and the separation between $1s$(K/K$^{\prime}$) and $2p$($\Gamma$) is about 0.3~eV in agreement with the separation of the experimentally measured binding energy differences (see below).

\section{STM/STS measurement parameters}
STM/STS measurements were performed using a Createc scanning probe microscope under ultrahigh vacuum ($<2\times10^{-10}$\,mbar) at low temperature ($\sim$6 K) with a Au terminated tip verified as metallic on a bare Au(111) substrate. All STS measurements were taken at constant height using a lock-in amplifier with a modulation frequency of 683\,Hz and amplitudes between 2\,mV  for point spectra and 10\,mV for maps.

\section{Experimental binding energies}

\begin{figure}[!h]
    \centering
    \includegraphics[width=0.5\textwidth]{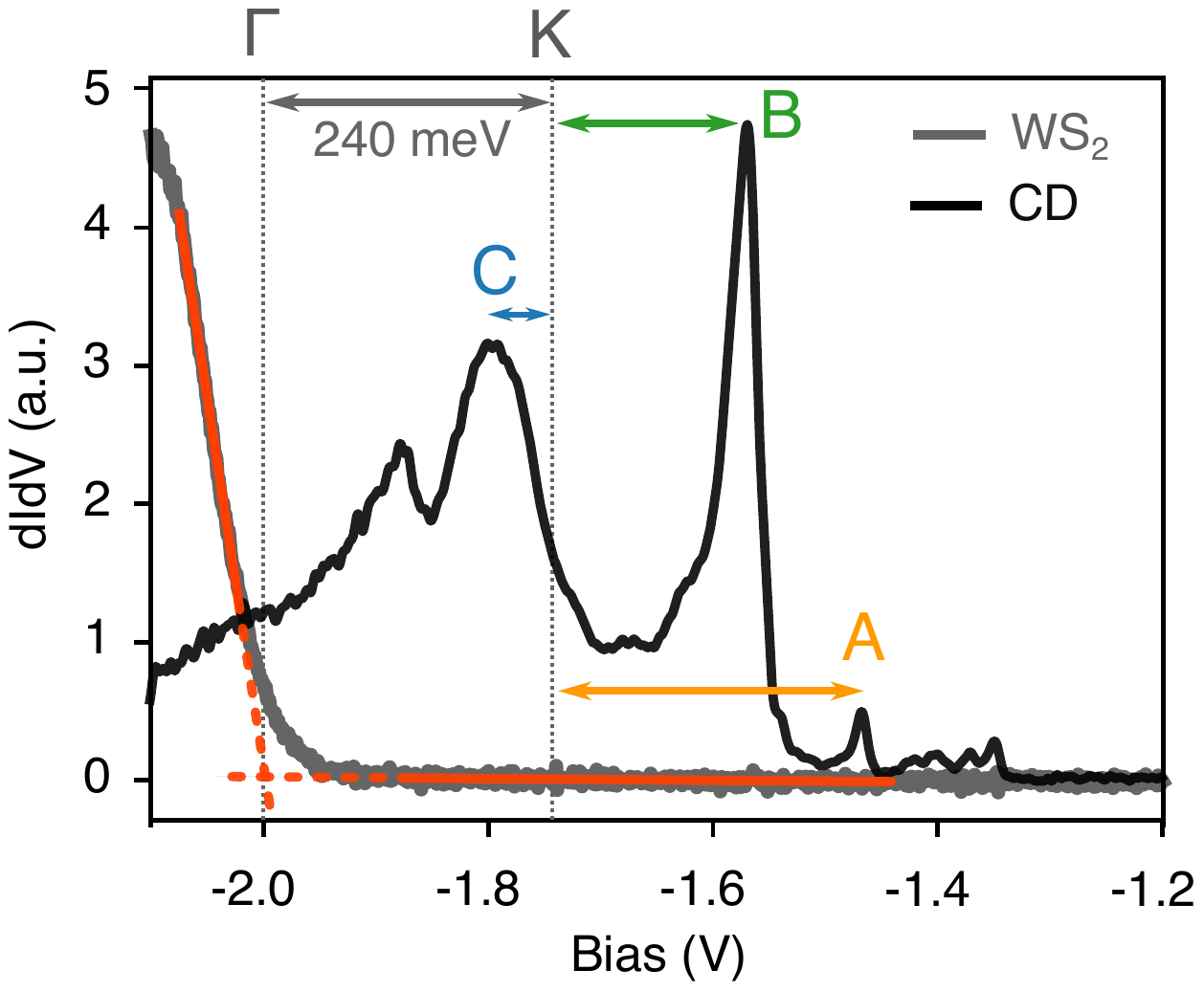}
    \caption{Determination of the experimental binding energies of states A, B and C. Red lines indicate linear fits used to extrapolate the $\Gamma$ band onset from the WS$_2$ spectrum (gray). Yellow, green, and blue arrows indicate the binding energy with respect to the K band of the A, B, and C resonance respectively.} 
    \label{fig:binding}
\end{figure}

The binding energies of electronic resonances A, B, and C were determined from the tunneling spectra. Since the K band onset is difficult to observe unless at very close tip-sample separations, the $\Gamma$ band was used as a reference. The $\Gamma$ onset was determined by linear extrapolation of zero crossing (see red lines in fig \ref{fig:binding}). The energy difference between the K band and the $\Gamma$ band of 240\,meV, as determined previously from ARPES measurements,~\cite{Kastl2017cvd} was used to determine the onset of the K band. Binding energies were calculated as the difference between resonance peaks and the valence band onset. The values are given in the table of the main manuscript. 
%and displayed below in table \ref{tab:binding}.

%\begin{table}[!h]
%\centering
%\label{tab:binding}
%\begin{tabular}{P{1cm}P{2cm}P{2cm}|P{2cm}P{2cm}}
%  & \multicolumn{2}{c}{Experiment} & \hspace{0.5cm}\multicolumn{2}{c}{Theory}\\
%  \cline{2-3} \cline{4-5}
% & K & $\Gamma$ &\hspace{0.5cm} K & $\Gamma$ \\ \hline
%\multicolumn{1}{l|}{A} & -175\,meV & -415\,meV &\hspace{0.5cm} -95\,meV & %-355\,meV\\ \hline
%\multicolumn{1}{l|}{B} & -95\,meV & -335\,meV &\hspace{0.5cm} -25\,meV & %-285\,meV  \\ \hline
%\multicolumn{1}{l|}{C} &+130\,meV & -110\,meV&\hspace{0.5cm} +180\,meV & %-80\,meV \\ 
%\end{tabular}
%\caption{Comparison of experimentally determined and theoretically calculated %energies of three observed resonances: A, B, and C. Values are listed with %respect ot the K and $\Gamma$ edge. Positive positive binding energies indicate %a higher energy of the respective state with respect to the band edge.}
%\end{table}

\section{Additional resonances observed in the band gap}

\begin{figure}[!h]
    \centering
    \includegraphics[width=0.5\textwidth]{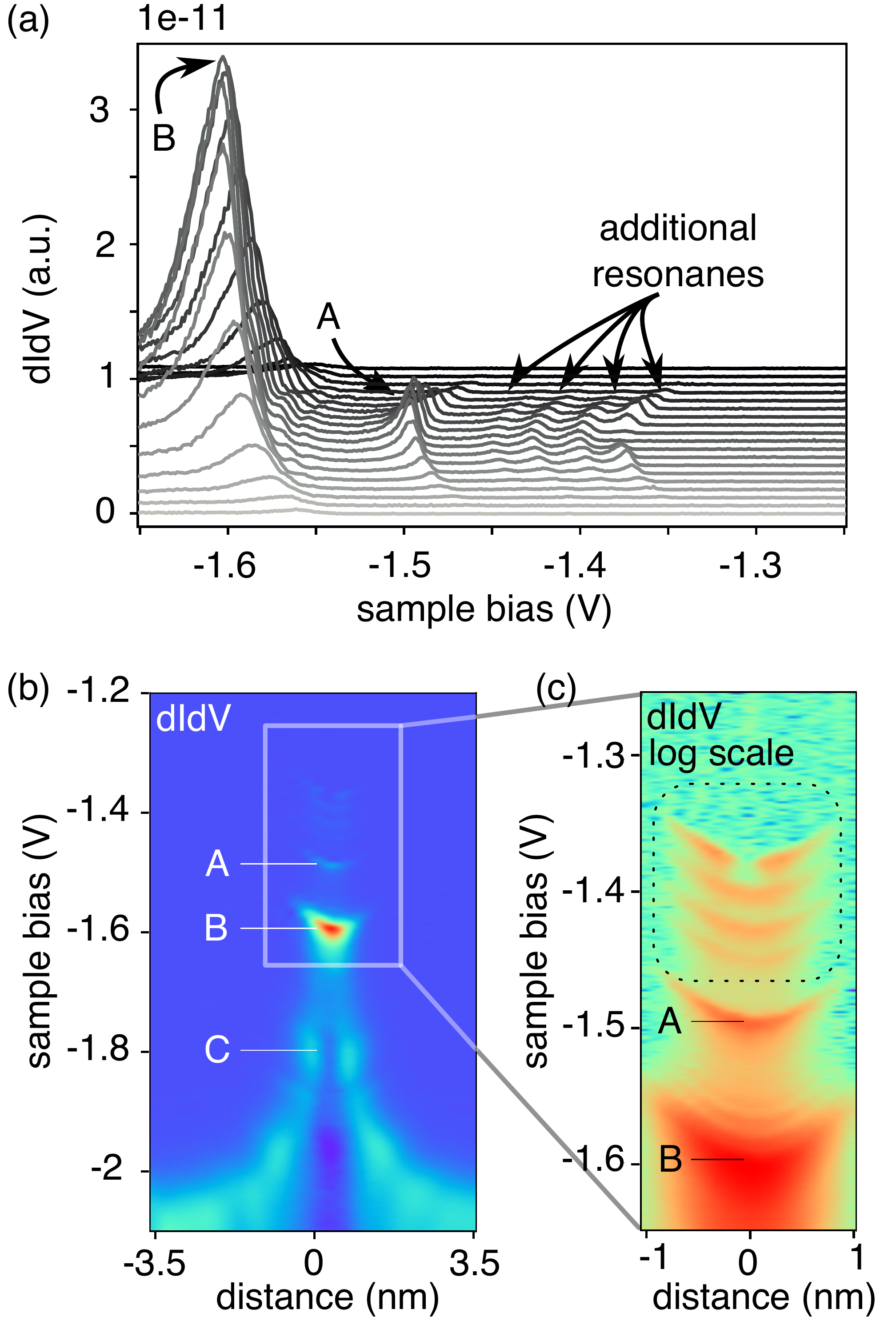}
    \caption{Scanning tunneling spectroscopy measurements of the charged defect showing the additional resonances above A. (a) Waterfall plot of constant height tunneling spectra taken along a line of a CD with states A and B labeled as well as the additional resonances observed between -1.35\,V and -1.45\,V. (b,c) Corresponding $\mathrm{d}I/\mathrm{d}V$ maps spatially resolving the additional states. Logarithmic color scale $\mathrm{d}I/\mathrm{d}V$ enhances the contrast of the states which are outlined with a black dotted box as shown in (c).}
    \label{fig:chem_staetes}
\end{figure}

In addition to the bound and resonance states A, B and C, there are four weak additional resonances observed just above state A, as shown in fig.~\ref{fig:chem_staetes}. The origin of these states is currently unclear.

\section{Simulated STM images}
To calculate the spatial STS signatures of the resonant impurity states, we note that their components exhibit very different tunnelling matrix elements. In particular, the large in-plane momentum of the continuum states at K results in a short out-of-plane decay length and small tunnelling matrix element. In contrast, the contributions to the resonant impurity states from the $\Gamma$ valley are expected to have a much larger tunnelling matrix element. As a first approximation, we can therefore construct the STS image of a resonant defect state simply by neglecting the contribution of the continuum states from the K valley.

\bibliography{ws2defect}